# Influence of Stoichiometry on the Optical and Electrical Properties of Chemical Vapor Deposition Derived MoS$_2$

In Soo Kim[1], Vinod K. Sangwan[1], Deep Jariwala[1], Joshua D. Wood[1], Spencer Park[1], Kan-Sheng Chen[1], Fengyuan Shi[1], Francisco Ruiz-Zepeda[5], Arturo Ponce[5], Miguel Jose-Yacaman[5], Vinayak P. Dravid[1,4], Tobin J. Marks[1,2], Mark C. Hersam[1,2,3,*], and Lincoln J. Lauhon[1,*]

[1]Department of Materials Science and Engineering, Northwestern University, Evanston, IL 60208 USA
[2]Department of Chemistry, Northwestern University, Evanston, IL 60208 USA
[3]Department of Medicine, Northwestern University, Evanston, IL 60208 USA
[4]International Institute of Nanotechnology, Northwestern University, Evanston, IL 60208 USA
[5]Department of Physics and Astronomy, University of Texas at San Antonio, San Antonio, TX 78249 USA

**Abstract**

Ultrathin transition metal dichalcogenides (TMDCs) of Mo and W show great potential for digital electronics and optoelectronic applications. Whereas early studies were limited to mechanically exfoliated flakes, the large-area synthesis of 2D TMDCs has now been realized by chemical vapor deposition (CVD) based on a sulfurization reaction. Since then, the optoelectronic properties of CVD grown monolayer MoS$_2$ have been heavily investigated, but the influence of stoichiometry on the electrical and optical properties has been largely overlooked. Here we systematically vary the stoichiometry of monolayer MoS$_2$ during CVD *via* controlled sulfurization and investigate the associated changes in photoluminescence and electrical properties. X-ray photoelectron spectroscopy is employed to measure relative variations in stoichiometry and the persistence of $MoO_x$ species. As $MoS_{2-\delta}$ is reduced (increasing $\delta$), the field-effect mobility of monolayer transistors increases while the photoluminescence yield becomes non-uniform. Devices fabricated from monolayers with the lowest sulfur content have negligible hysteresis and a threshold voltage of ~0 V. We conclude that the electrical and optical properties of monolayer $MoS_2$ crystals can be tuned *via* stoichiometry engineering to meet the requirements of various applications.

\* Correspondence to: lauhon@northwestern.edu, m-hersam@northwestern.edu

The physical properties of electronic materials can be usefully controlled by tuning stoichiometry and doping. In conventional semiconductors like silicon, doping improves the electrical conductivity by increasing carrier concentration without significantly degrading carrier mobility.[1] However, doping of reduced dimensionality semiconductors without degrading mobility has proven challenging due to confined channel dimensions and increased scattering cross-section of carriers, as seen in reduced dimensionality carbon allotropes such as carbon nanotubes and graphene.[2] Similar challenges arise for the layered transition metal dichalcogenides (TMDCs) of Mo and W. Mono- and few layer TMDCs exhibit bandgaps that can be tuned by thickness, strain, and composition, making them promising 2D semiconductors complementary to graphene and hexagonal boron nitride in various optoelectronic applications.[3] The electrical and optical properties of TMDCs are particularly sensitive to surrounding interfaces,[4-8] as well as intrinsic and extrinsic defects such as chalcogen vacancies[9,10] and dopants,[11-13] respectively. Consequently, defect engineering through controlled synthesis and post synthesis processing provides a facile means to tune physical properties.

Several CVD-based methods have been developed to synthesize $MoS_2$ over large areas, including sulfurization of e-beam evaporated thin Mo metal films,[14] decomposition of ammonium thiomolybdate $((NH4)_2MoS_4)$,[14] direct vapor phase deposition of $MoS_2$ powder,[15] and sulfurization of molybdenum trioxide $(MoO_3)$ using elemental sulfur.[16-18] Among these methods, the sulfurization of $MoO_3$ has been the most widely adopted. Because varying degrees of sulfurization during growth may lead to varying concentrations of both intentional and unintentional point defects, such as vacancies and dopants, it is important to establish the influence of materials processing conditions on physical properties. For example, the field-effect mobility of both mechanically exfoliated and as-synthesized CVD MoS$_2$ monolayer crystals, hereafter referred to as "flakes", are an order of magnitude lower than the intrinsic limits.[19, 20] Variations in carrier mobilities and charge transport mechanisms in CVD MoS$_2$ have been variously attributed to trapped charges at sulfur vacancies,[14] trapped charges at the interface of MoS$_2$ and oxide dielectrics,[21] extrinsic disorder from adsorbates,[20, 22] grain boundaries,[17] and other defects within the films.[3, 18, 20, 23] However, charge transport in CVD MoS$_2$ has not been investigated in concert with systematic variations in stoichiometry.

In this work, we correlate stoichiometry with the electrical properties of CVD synthesized monolayer MoS$_2$ flakes prepared under varying degrees of MoO$_3$ sulfurization. X-ray photoelectron spectroscopy shows that an increase in the degree of sulfurization leads to an improvement in the relative stoichiometry and a decrease in the amount of MoO$_3$/MoO$_x$ within the region probed. Approaching stoichiometric MoS$_2$, the photoluminescence becomes spatially more uniform while the electrical transport becomes more resistive with large hysteresis in gate bias sweeps. Surprisingly, the least stoichiometric samples show up to an order of magnitude higher field-effect mobilities, negligible hysteresis, and threshold voltages close to 0.0 V. An inverse correlation between stoichiometry and standard transistor metrics is established among a set of three different growth conditions, demonstrating the potential to optimize stoichiometry for various device applications.

**Results and Discussion**

To investigate the effects of stoichiometry on the optical and electrical properties of monolayer $MoS_2$ flakes, three groups of samples were prepared (Table 1) by modifying previously described growth procedures.[16-18] Briefly, $MoO_3$ and S powder sources were evaporated at 800 °C and 150 °C, respectively. The sulfur source was loaded sequentially in a total of four steps to prevent premature sulfurization of the $MoO_3$ source and to ensure sufficient sulfurization throughout the growth (see details in Methods). The degree ofسulfurization was controlled by varying the time of exposure to sulfur vapor and the temperature of sulfur source as tabulated in Table 1.

Table 1. Variation in growth conditions for intentional modification of stoichiometry in $MoS_2$ flakes.

|  | Group A | Group B | Group C |
| --- | --- | --- | --- |
| Exposure to Sulfur Vapor (mins) | 10 | 10 | 3 |
| Temperature of Sulfur (°C) | 170 | 150 | 150 |

Individual monolayer $MoS_2$ flakes of all the samples grown on $SiO_2$/Si (300 nm) are readily identified by the optical contrast arising from thin film interference (Figure 1a). Atomic force microscopy (AFM) was used to determine the absolute thickness of individual flakes as well as to visualize thickness variations and grain boundaries within flakes. A thickness of ~5 Å was extracted from the height profile of the AFM topography image of an individual $MoS_2$ flake (Figure 1b), in line with previous reports.[4, 24-26] Interestingly, the grain boundaries in mono- and bilayer regions exhibit a sharp contrast in the AFM phase image (Figure 1c and Supporting Information). Brightness and contrast of the bilayer region was adjusted to distinguish mono- and bilayer regions. Similar grain boundaries in MoS$_2$ flakes have been visualized with transmission electron microscopy (TEM); however, the AFM imaging shown here enables rapid and non-destructive imaging of atomically thin grain boundaries. The local atomic structure of a flake consisting of both mono- and bilayer MoS$_2$ was next examined by aberration (C$_s$) corrected scanning transmission electron microscopy (STEM) in a JEOL ARM 200F instrument operated at 80 kV. The crystal is flat and a six-fold symmetry was confirmed directly from the STEM image with lattice spacing of 3.1 ± 1 Å (Figure 1d). In the high angular dark field (HAADF) image, the contrast is proportional to the square of the atomic number, and atomic positions of molybdenum and sulfur atoms can be distinguished; Mo atoms show higher contrast than S atoms. This enables identification of the stacking sequence in MoS$_2$ bilayers. The Raman spectrum in Figure 1e of monolayer MoS$_2$ exhibits characteristic E$^1_{2g}$ and A$_{1g}$ modes associated with in-plane and out-of plane vibrations. The two peaks appear at ~383.8 and 404.4 cm$^{-1}$, resulting in the frequency difference of ~20.6 cm$^{-1}$, consistent with the values reported in literature for CVD monolayer MoS$_2$ flakes. A strong luminescence band located at 678 nm (excitation wavelength at 532 nm) shown in Figure 1f corresponds to the A direct excitonic transition,[27] indicative of a monolayer MoS$_2$ flake with a direct 1.83 eV band gap.

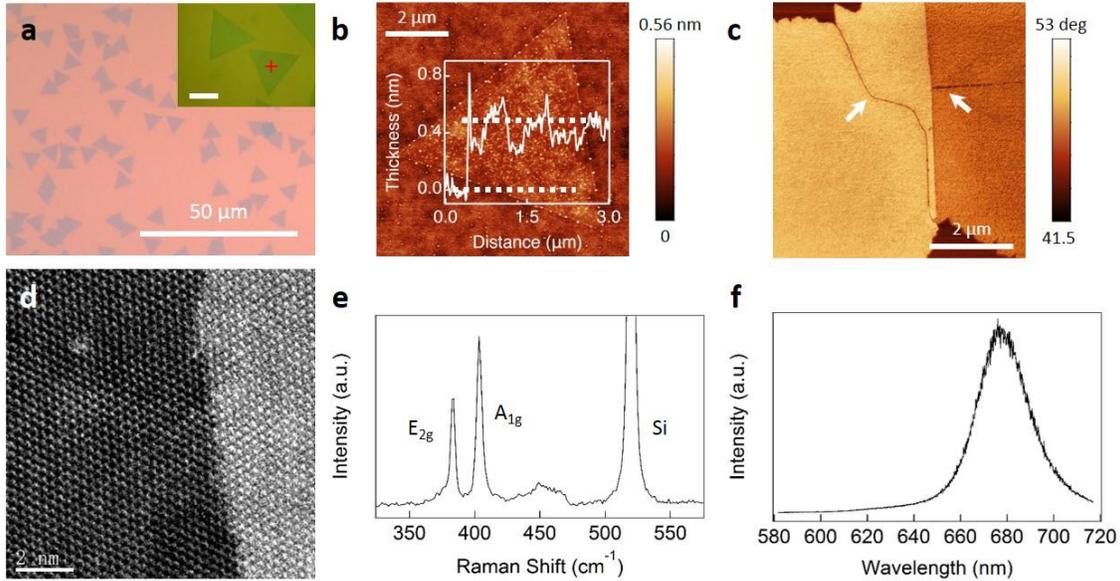

**Figure 1.** a) Optical image of monolayer $MoS_2$ flakes. Inset is a higher magnification image. The cross hair indicates where Raman and photoluminescence spectra were acquired. Scale bar 10 μm. b) AFM topography image of isolated monolayer $MoS_2$ and corresponding height profile (inset). c) AFM phase image of a $MoS_2$ flake consisting of mono- and bilayer regions with grain boundaries marked by white arrows. Bilayer region is distinguished by the dark contrast. d) High angle annular dark field (HAADF) STEM image resolving the atomic structure. The image was recorded using a convergence angle of 25 mrad with a probe size of about 0.09 nm. e), f) Representative Raman and photoluminescence spectra of monolayer $MoS_2$.

**Photoluminescence and Raman Microspectroscopy**

Representative variations in the photoluminescence and Raman spectra within flakes and between sample groups are shown in Figure 2. Photoluminescence emission and Raman $E^1_{2g}$ mode scattering maps were constructed by integrating the respective emission and scattering intensities. Multiple samples were investigated from each of the groups to establish a representative data set. Group A samples (highest sulfur exposure) exhibit spatially uniform emission intensity and strong emission peaked at ~678 nm (Figures 2a,d). The peak emission wavelength and full width at half maximum (FWHM) vary less than 0.4 nm across the flake. In contrast, the group C sample (least sulfur exposure) exhibits in-plane variations in emission intensity (Figure 2c,f), with the emission intensity in the central region ~3 times higher than that of the edge regions, but only ~35 % of that of the group A sample. Furthermore, the center emission is slightly red-shifted, although no strong correlation is observed between the emission intensity and wavelength. The full width at half maximum (FWHM) of the emission peaks is comparable between sample groups. The photoluminescence intensity of the group B samples is comparable to that of the group A samples, but the intensity decreases moving towards the center of the flake (Figure 2b,d).

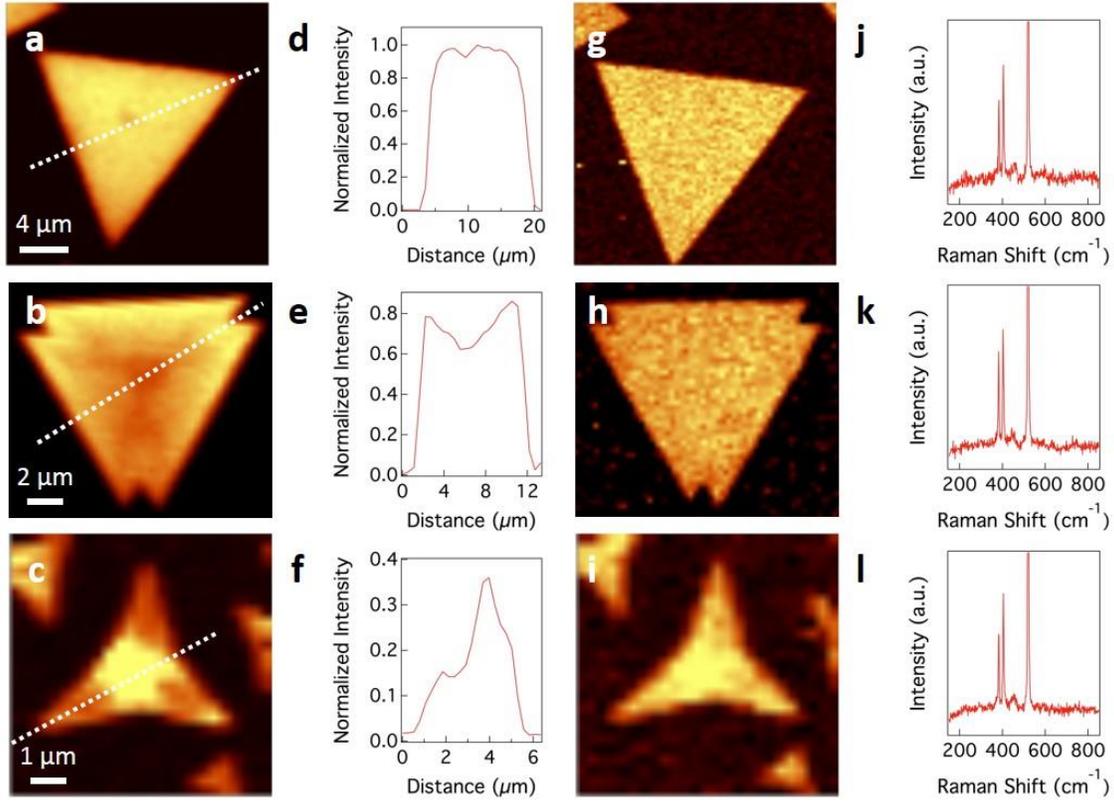

**Figure 2.** a), b), c) Integrated photoluminescence intensity maps for samples from group A, B, and C, respectively. d), e), f) Normalized intensity line profiles extracted from photoluminescence maps. g), h), i) Corresponding Raman maps constructed by integrating $E^1_{2g}$ mode. j), k), l) Representative Raman spectra for the three different growth conditions.

Many factors may influence photoluminescence characteristics including defects (vacancies, grain boundaries), strain, and electrostatic doping.[17, 28] Raman spectroscopy was used to rule out strain and electrostatic doping as primary causes of the variations in emission intensity; the frequency of $E^1_{2g}$ vibrational mode is sensitive to strain,[29] whereas the frequency of $A_{1g}$ vibrational mode is sensitive to electrostatic doping.[30] Raman maps of the $E^1_{2g}$ peak center of mass are quite uniform (Figure 2g,h,i), and are comparable to similar maps of the $A_{1g}$ mode. Integrated peak intensities and FWHM are also uniform, indicating that strain and electrostatic doping are negligible. We do not observe Raman signatures from oxide phases ($MoO_3$ and/or $MoO_x$) that would indicate the presence of residual oxygen associated with incomplete sulfurization (Figure 2 j,k,l). The possible contribution of sulfur vacancies to the non-uniform photoluminescence characteristics is discussed below.

**Analysis of Stoichiometry by XPS**

X-ray photoelectron spectroscopy (XPS) analyses of the three groups of samples establish systematic variations in stoichiometry associated with the distinct processing conditions (Figure 3). To compensate for sample charging, all spectra were charge corrected against the C 1s adventitious carbon peak at 284.8 eV. In Mo 3d spectra of the group A (most stoichiometric) sample, both $Mo^{4+}$ and $Mo^{6+}$ doublets are observed. $Mo^{4+}$ $3d_{5/2}$ and $3d_{3/2}$ doublet peaks appear at ~229.8 and ~233.0 eV with FWHM of ~ 1.23 eV (Fig. 3a). The doublet peaks of $Mo^{4+}$ $3d_{5/2}$ and $3d_{3/2}$ were deconvoluted into two components to obtain a good fit. The first set of components, located at 229.5 and 232.7 eV, corresponds to stoichiometric intrinsic $MoS_2$ (i-$MoS_2$). The second set of components located at slightly lower binding energies of 229.2 and 232.2 eV correspond to defective/sub-stoichiometric $MoS_2$ (d-$MoS_2$) with sulfur vacancies. The third component located under $Mo^{4+}$ $3d_{3/2}$ doublet and the broad peak (232.6 and ~236.0 eV) represents $Mo^{6+}$ $3d_{5/2}$ and $3d_{3/2}$ doublets of $MoO_3$[31, 32] or $MoO_x$ suboxides,[33] respectively. While we did not find evidence of $MoO_3/MoO_x$ within individual flakes from Raman mapping, the presence of $MoO_3/MoO_x$ in XPS likely arises from the large probe size (~400 μm diameter), which results in sampling of extended regions enclosing thicker regions of $MoS_2$ that may have incompletely sulfurized. We further note that the degree of inhomogeneity across the area sampled by XPS likely exceeds the degree of inhomogeneity within a single flake.

As expected, the presumed substoichiometric group C sample exhibits noticeable changes in its XPS spectra. First, shifts to lower binding energies were observed for the $Mo^{4+}$ $3d_{5/2}$ and $3d_{3/2}$ doublet peaks, which reflect reduction of $MoS_2$ consistent with the presence of sulfur vacancies.[34, 35] A similar behavior was observed previously through preferential sputtering of sulfur from $MoS_2$.[34, 35] When the doublets are decomposed using the same components used for the group A sample, the contribution of the intrinsic $MoS_2$ (i-$MoS_2$) decreases, whereas the defective $MoS_2$ (d-$MoS_2$) component increases. Second, the $MoO_3/MoO_x$ component is larger in the group C sample, indicating the presence of non-stoichiometric $MoS_2$ from incomplete sulfurization of $MoO_3/MoO_x$, and the increase in FWHM to ~1.27 eV indicates a higher degree of disorder in the sub-stoichiometric group C sample. Finally, the decreased integrated area of the S 2s peak with respect to the Mo 3d peaks indicates a reduction in the total amount of sulfur as compared with the group A sample.

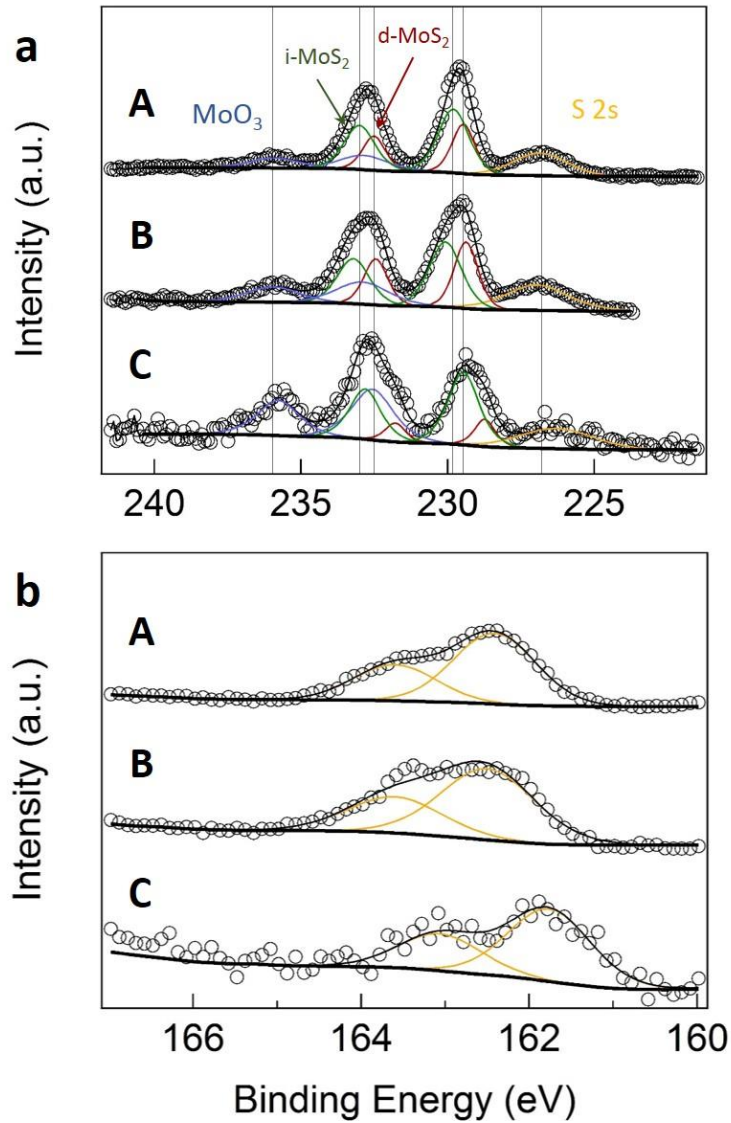

**Figure 3.** X-ray photoelectron spectra (XPS) for different CVD $MoS_2$ growth conditions. a) Mo 3d and b) S 2p core level spectra for CVD $MoS_2$ flakes on $SiO_2$/Si. Chemical contributions from $MoO_3$, intrinsic $MoS_2$ (i-$MoS_2$), and defective $MoS_2$ (d-$MoS_2$) are shown.

**Transport measurements**

To investigate the effects of stoichiometry on the transport properties of CVD $MoS_2$, two- and four-terminal devices were fabricated on individual, isolated $MoS_2$ flakes without grain boundaries using 2 nm Ti/70 nm Au metal contacts. An optical image of a representative four-terminal MoS$_2$ device (group A) in Figure 4a shows the trapezoidal channel for van der Pauw measurements. All devices were measured at room temperature at pressures of ~2 × 10$^{-5}$ Torr. The majority of devices were fabricated and measured in a two-terminal geometry, while four-terminal measurements of a selected sub-set of devices were conducted to assess the influence of contact resistance on mobility estimates and transistor characteristics (Supporting Table 1). Due to the large impedance in the sub-threshold regime (see Methods), four-terminal van der Pauw measurements were conducted in the low impedance accumulation regime ($V_g > V_{th}$).

In general, as-fabricated group A devices had the lowest field-effect mobilities and largest hysteresis among samples considered here (Supporting Table 1). Figure 4b shows transfer characteristics (channel conductance *versus* gate voltage) of one device (114-S1, see Supporting Table 1) after conducting a current-annealing procedure, which was found to improve the device characteristics as discussed further below and in Supporting Information Section 2. The field-effect mobility (μ$_{FE}$) was calculated from the transfer curves based on the relation:

$$\mu_{FE} = \frac{dI_d}{dV_g}\left[\frac{L}{WC_iV_d}\right],$$

where $I_d$, $V_g$, and $V_d$ are the drain current, gate voltage, and drain voltage, respectively; $L$ is the length and $W$ is the width of the channel; and ~11 nF/cm$^2$ is assumed for the area normalized capacitance, $C_i$, of the 300 nm thick SiO$_2$. The field-effect mobility of six *as-fabricated* group A devices are in the range of 5 × 10$^{-3}$ to 1.5 cm$^2$/Vs while the average value (0.48 cm$^2$/Vs) is comparable to values reported in literature.[16, 17]

Compared to exfoliated $MoS_2$, Group A devices show larger hysteresis,[5] larger threshold voltages (average $V_{th}$ = 40.8 V), and lower $I_{ON}/I_{OFF}$ ratios (~$10^4$).[5] Furthermore, non-linear output characteristics at low biases in as-fabricated devices suggested the presence of a Schottky barrier at the contacts (Supporting Information Section 2). As a result, highly resistive as-fabricated devices could not be probed *via* four-terminal van der Pauw measurements. However, the transport characteristics of group A devices could be improved *via* current-annealing in vacuum as discussed in Supporting Information Section 2. Following annealing, the average field-effect mobility of the more resistive devices was increased by up to two orders of magnitude (*e.g.* from $5 \times 10^{-3}$ to 0.16 $cm^2$/Vs for sample 114-S1 in Supporting Table 1). Although current annealing also improves the linearity of *I-V* curves, they remain less linear than those of exfoliated $MoS_2$ flakes[5] and the sulfur-deficient CVD $MoS_2$ samples discussed below (Supporting Fig S1c, S2a). The relative enhancement in device performance by current-annealing, which presumably involves Joule heating, is consistent with a recent report on thermal annealing of $MoS_2$.[36]

Group B samples with a lower sulfur content (Fig. 3) show slightly higher field-effect mobilities (0.2 – 3.8 $cm^2$/Vs), lower threshold voltages, and reduced hysteresis (Supporting Fig. S3, Table 1) compared to group A samples. The average field-effect mobility, threshold voltage, and $I_{ON}/I_{OFF}$ ratio values of six group B devices are 2.66 $cm^2$/Vs, 34.8 V, and ~$10^5$, respectively. Group C samples, which are the least stoichiometric, showed a continued trend in electrical characteristics. A representative two-terminal device exhibits linear output characteristics at low biases (Figure 4c), suggesting that the contacts are improved compared to group A and B devices.[4, 37] Significantly improved transport characteristics of the group C devices are also evident in the linear and semi-log transfer curves of the same device for both two- and four-terminal measurements (Fig. 4d). First, compared with groups A and B, the field-effect mobilities of group C devices are larger by approximately by an order of magnitude, in the range of 12 – 21 cm$^2$/Vs with an average mobility of 15.3 cm$^2$/Vs. This is the highest value reported for as-synthesized unencapsulated CVD monolayer MoS$_2$ flakes without current or thermal annealing. Furthermore, the threshold voltage shifted towards negative bias to an average of 12.3 V, which is indicative of increased *n*-type doping. The $I_{ON}/I_{OFF}$ ratio was also improved to ~10$^6$ without noticeable hysteresis, consistent with the generally better transistor performance of the group C samples.

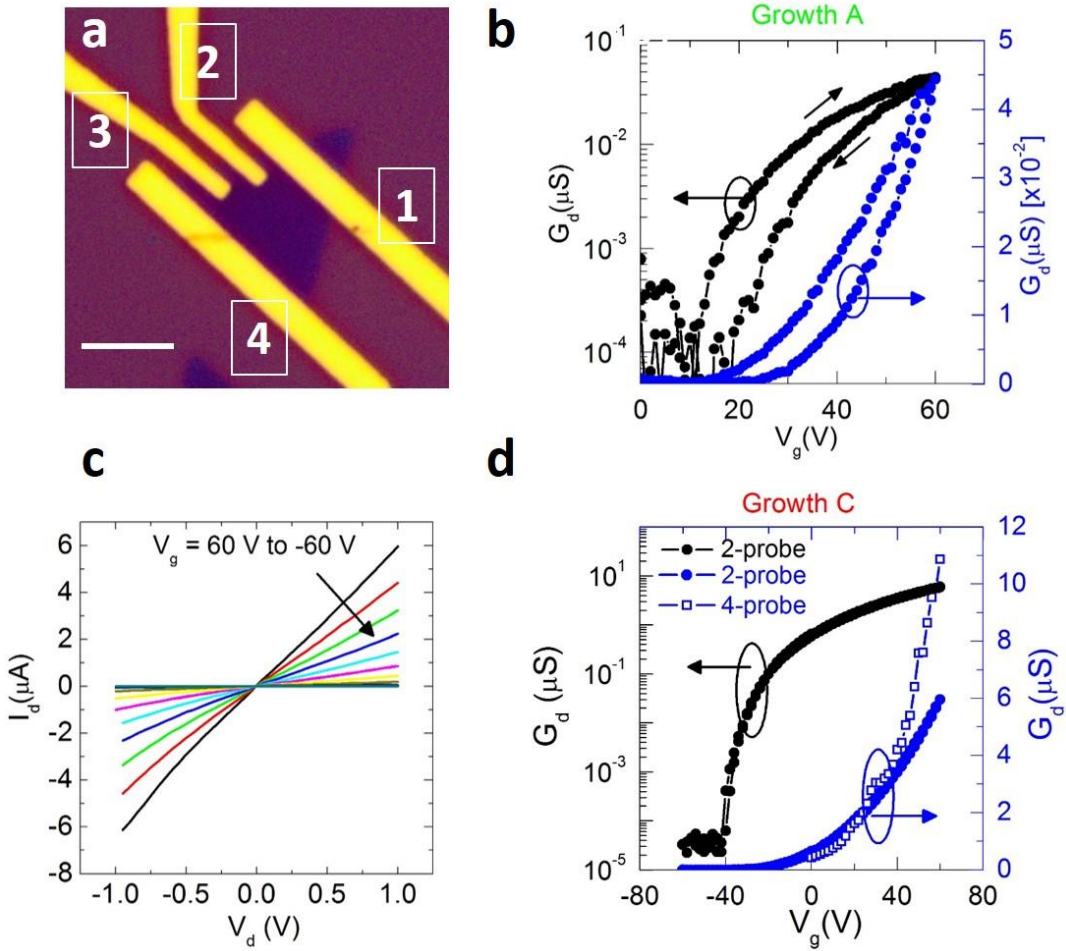

**Figure 4.** a) Optical image of a 4-probe MoS$_2$ device (growth A). Scale bar is 4 µm. b) Two-terminal transfer characteristics of the device measured between electrode "1" and "4" at $V_d$ = 50 mV in linear as well as semi-log plots. c) Output $I_D$-$V_D$ characteristics of a growth C device at $V_G$ = 60 V to −60 V. d) Conductance ($G_d$) *versus* gate voltage ($V_g$) of the same device. Both two-terminal and four-terminal conductance are shown.

A comparison of the transfer characteristics of devices from the three growth conditions associates an increase in the average field-effect mobility and decrease in the threshold voltage with decreasing S content (Figure 5). As a proxy for direct composition measurements, we plot the ratio of the integrated S and MoO$_3$/MoO$_x$ peaks for the three growth conditions in Figure 5b, normalized against the relative sensitivity factors for the respective core levels. Considering Figures 4 and 5 together, we conclude that *n*-type conductivity, field-effect mobility, doping, and $I_{ON}$/$I_{OFF}$ ratio increase with decreasing stoichiometry. The increase in electron doping can be explained by an increasing concentration of sulfur vacancies acting as electron donors, but it is perhaps surprising that the field-effect mobility also increases.[9, 10] Below we examine whether the PL data are also consistent with the proposed variation in sulfur vacancy concentration, and then consider possible explanations for the trends in field effect mobility.

In general, structural defects reduce emission intensity when they generate mid-gap states that provide non-radiative recombination pathways.[28] In a previous study of monolayer MoS$_2$, photoluminescence quenching (enhancement) near mirror (tilt) boundaries,[17] was attributed to increased (decreased) free carrier density near such boundaries, as would be produced by sulfur vacancies. If sulfur vacancies[12, 35] indeed cause quenching, one might expect to observe variations in photoluminescence intensity between samples of differing stoichiometry, and even within samples that are sulfur deficient. The group A, B, and C samples exhibit expected variations with decreasing sulfur stoichiometry. First, the samples have different morphologies that show the same trend with decreasing sulfur exposure as was previously reported by the Ajayan group.[16] Additionally, group C samples show significant variations in PL intensity within individual flakes, consistent with an increased sulfur vacancy concentration.

Structural defects and stoichiometry also have a strong influence on charge transport. Recent theoretical and experimental work has established that carrier mobilities and charge transport mechanisms in 2D $MoS_2$ depend strongly on carrier densities.[5, 36, 38-40] Specifically, single-layer $MoS_2$ undergoes a metal to insulator transition as a function of carrier density. In the metallic state, the Fermi level is close to or above the mobility edge while in the insulating state the Fermi level is deep inside the band gap within a tail of localized states that gives rise to hopping transport.[5, 36, 38] Indeed, capacitance-voltage and transport measurements by Zhu *et al.*[21] on CVD $MoS_2$ devices with similar transfer characteristics to those described here identified the influence of localized (trap) states arising from structural defects. However, it has been established that S vacancies are sufficiently shallow to act as electron donors in $MoS_2$.[41, 42] For the particular distribution of defects present in these CVD grown films, it appears that the electron donation associated with sulfur vacancies makes a net positive contribution to the conductivity *via* both an increase in electron concentration and a related increase in field-effect mobility.

We note that this counterintuitive result would not be expected for the introduction of sulfur vacancies in perfect materials with band transport.

Furthermore, the trends in mobility and carrier concentration cannot be readily explained in terms of variations in oxygen content. First, one would expect a *decrease* in conductivity upon formation or persistence of disordered insulating $MoO_3$ domains.[43] Second, replacement of residual oxygen atoms with sulfur atoms should actually increase the mobility, and replacement of residual oxygen atoms with vacancies could also improve the transconductance, from which the field-effect mobility is derived. Calculations of the scattering rates produced by substitutional oxygen and sulfur vacancies, as well as atomic level confirmation of the relative concentrations of each type of defect, would be useful in developing a more complete understanding of the factors that control mobility in CVD grown material.

We have shown here that the tuning of stoichiometry provides a useful degree of control over important device parameters, but it is important to consider what factors limit the performance of our materials in the context of recent work.[17, 18, 21] Metal contacts do not have a dominant effect since field-effect mobilities extracted from 2-terminal and 4-terminal measurements of group A as well as group C devices are comparable. While the mobility of ~20 $cm^2$/Vs is the highest value reported for as-synthesized unencapsulated CVD monolayer $MoS_2$ crystals at room temperature, vacuum thermal annealing of CVD grown material has been shown to increase the field-effect mobility up to ~100 $cm^2$/Vs at room temperature,[20] indicating room for further improvement. As noted above, capacitance-voltage and transport measurements by Zhu *et al.*[21] on CVD $MoS_2$ devices with similar transfer characteristics to those described here identified the influence of localized (trap) states arising from structural defects. Trap states below the mobility edge reduce the transconductance even beyond the threshold voltage and likely lead to an underestimate of the free-carrier field-effect mobility of the present CVD $MoS_2$ transistors.

**Conclusions**

We summarize with three observations based on this work and the recent works discussed above. First, stoichiometry variations contribute to unexpected variations in carrier mobilities. Second, since our stoichiometric (non-stoichiometric) flakes exhibit nominally better optical (electrical) properties, the optimal synthetic conditions cannot necessarily be deduced by electrical or optical measurements alone. Third, given the recent increases in mobility achieved by vacuum annealing,[20] a combined investigation of CVD growth conditions and post-growth annealing in vacuum and other environments is needed to assist in the identification of native defects and work towards their minimization in $MoS_2$ and other TMDCs. Atomically resolved studies involving annular dark field (ADF) STEM or scanning tunneling microscopy (STM) may prove useful in this regard.

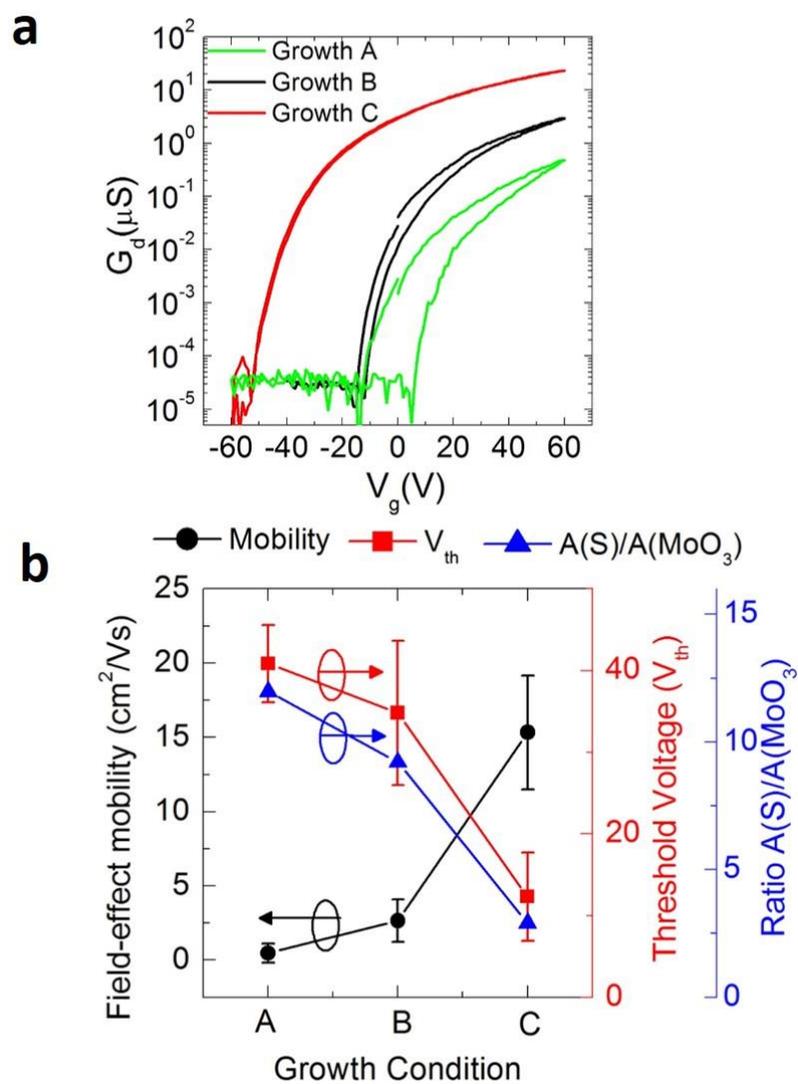

**Figure 5.** a) $G_d$ *versus* $V_g$ ($V_d = 0.5$ V) of typical devices from growth A, B, and C. b) Comparison of field-effect mobility, threshold voltage ($V_{th}$), and the normalized ratio of the S 2p and MoO$_3$ 3d XPS core level areas with growth condition.

**Methods**

*CVD Synthesis of Monolayer MoS$_2$ Flakes*

Si wafers with 300 nm of thermally grown SiO$_2$ were cleaned in an O$_2$ plasma and used as growth substrates. Alumina boats containing solid MoO$_3$ (99 %, Sigma-Aldrich) and S (99.95 %, Sigma-Aldrich) powders were used as Mo and S precursors by placing in a 25 mm quartz tube in temperature zones of 800 °C and 150 °C, respectively. Growths were carried out at 150 Torr with He as the carrier gas (20 sccm). During the temperature ramp, the sulfur source (initially 3.85 inches away from the end of the tube) was loaded stepwise to 4.55, 4.95, and 5.25 inches away from the tube end when the temperature of the MoO$_3$ source reached 650, 725, and 800 °C, respectively. The growth was carried out for 10 min prior to cooling under a He flow.

*X-ray Photoelectron Spectroscopy*

XPS was conducted with a Thermo Scientific ESCA Lab 250Xi XPS with a monochromatic Kα-Al X-ray line. The probe size was ~ 400 μm in diameter, nominally elliptical in cross section. A charge neutralization flood gun (Ar$^+$ ions) was used to compensate for local electrostatic fields on the MoS$_2$/SiO$_2$/Si samples. All elements within surveys were fitted with Shirley backgrounds. Core level spectra were charge corrected against fitted adventitious carbon at 284.8 eV. Sub-peaks were fit using modified Shirley backgrounds and floating Gaussian-Lorentzian (GL) mixing. All sub-peaks shared the same amount of GL character, but that amount was kept as a fitting parameter. Subpeaks were fit such that their full-width at half maximum (FWHM) values were less than 3 eV. Core level spectra were collected at a pass energy of 100 eV and a dwell time of 100 ms.

*Device Fabrication and Measurement*

Field-effect transistors were fabricated on thermal oxide Si substrates (300 nm SiO$_2$) using standard e-beam lithography and lift-off processes. Metal contacts (2 nm Ti/70 nm Au) were thermally evaporated. After lift-off, devices were cleaned in N-Methyl-2-pyrrolidone and deionized water to remove processing residues, which was observed to decreased hysteresis in MoS$_2$ transistors. All measurements were conducted in vacuum (pressure $< 2 \times 10^{-5}$ Torr) in a LakeShore CRX 4K probe station using Keithley source-meters and home-made LabVIEW programs. Four-terminal measurements were conducted by passing 10 nA current between electrode "1" and "4" in Fig. 4a using Keithley 6430 source-meter and employing a remote preamplifier. Four-probe measurements of the devices in the sub-threshold regime (Fig. 4d) were not possible due to limits of input impedance of the instrument. Conductance *versus* gate voltage measurements of 4-terminal devices in the linear regime ($V > V_{th}$) were used to calculate the field-effect mobility.

**Acknowledgments**

This work was supported by the Materials Research Science and Engineering Center (MRSEC) of Northwestern University (National Science Foundation Grant DMR-1121262), NSF PREM Grant DMR-0934218, the Office of Naval Research (Grant N00014-14-1-0669), and the Keck Foundation. The microscopy work was supported by grants from the National Center for Research Resources (5 G12RR013646-12) and the National Institute on Minority Health and Health Disparities (G12MD007591) from the National Institutes of Health.

**Supporting Information Available**

AFM phase images of grain boundaries in MoS$_2$; Effects of current annealing on MoS$_2$ devices; Summarized electrical properties of MoS$_2$ devices. This material is available free of charge *via* the internet at http://pubs.acs.org.

**References**

1. Sze, S. M.; Ng, K. K. *Physics of Semiconductor Devices*; 3rd ed.; Wiley: New Jersey, 2007.

2. Jariwala, D.; Sangwan, V. K.; Lauhon, L. J.; Marks, T. J.; Hersam, M. C. Carbon Nanomaterials for Electronics, Optoelectronics, Photovoltaics, and Sensing. *Chem. Soc. Rev.* **2013**, 42, 2824-2860.

3. Jariwala, D.; Sangwan, V. K.; Lauhon, L. J.; Marks, T. J.; Hersam, M. C. Emerging Device Applications for Semiconducting Two-Dimensional Transition Metal Dichalcogenides. *ACS Nano* **2014**, 8, 1102–1120.

4. Radisavljevic, B.; Radenovic, A.; Brivio, J.; Giacometti, V.; Kis, A. Single-Layer $MoS_2$ Transistors. *Nat. Nanotechnol.* **2011**, 6, 147-150.

5. Jariwala, D.; Sangwan, V. K.; Late, D. J.; Johns, J. E.; Dravid, V. P.; Marks, T. J.; Lauhon, L. J.; Hersam, M. C. Band-Like Transport in High Mobility Unencapsulated Single-Layer $MoS_2$ Transistors. *Appl. Phys. Lett.* **2013**, 102, 173107.

6. Late, D. J.; Huang, Y.-K.; Liu, B.; Acharya, J.; Shirodkar, S. N.; Luo, J.; Yan, A.; Charles, D.; Waghmare, U. V.; Dravid, V. P.*, et al.* Sensing Behavior of Atomically Thin-Layered $MoS_2$ Transistors. *ACS Nano* **2013**, 7, 4879-4891.

7. Perkins, F. K.; Friedman, A. L.; Cobas, E.; Campbell, P. M.; Jernigan, G. G.; Jonker, B. T. Chemical Vapor Sensing with Monolayer $MoS_2$. *Nano Lett.* **2013**, 13, 668-673.

8. Sangwan, V. K.; Arnold, H. N.; Jariwala, D.; Marks, T. J.; Lauhon, L. J.; Hersam, M. C. Low-Frequency Electronic Noise in Single-Layer $MoS_2$ Transistors. *Nano Lett.* **2013**, 13, 4351-4355.

9. Park, J. B.; France, C. B.; Parkinson, B. A. Scanning Tunneling Microscopy Investigation of Nanostructures Produced by $Ar^+$ and $He^+$ Bombardment of $MoS_2$ Surfaces. *J. Vac. Sci. Technol. B* **2005**, 23, 1532-1542.

10. Sengoku, N.; Ogawa, K. Investigations of Electronic Structures of Defects Introduced by Ar Ion Bombardments on $MoS_2$ by Scanning Tunneling Microscopy. *Jap. J. Appl. Phys.* **1995**, 34, 3363-3367.

11. Mann, J.; Ma, Q.; Odenthal, P. M.; Isarraraz, M.; Le, D.; Preciado, E.; Barroso, D.; Yamaguchi, K.; von Son Palacio, G.; Nguyen, A.*, et al.* 2-Dimensional Transition Metal Dichalcogenides with Tunable Direct Band Gaps: $MoS_{2(1-x)}Se_{2x}$ Monolayers. *Adv. Mater.* **2014**, 26, 1399-1404.

12. Ma, Q.; Isarraraz, M.; Wang, C. S.; Preciado, E.; Klee, V.; Bobek, S.; Yamaguchi, K.; Li, E.; Odenthal, P. M.; Nguyen, A., *et al.* Post-Growth Tuning of the Bandgap of Single-Layer Molybdenum Disulfide Films by Sulfur/Selenium Exchange. *ACS Nano* **2014**, 8, 4672-4677.

13. Feng, Q.; Zhu, Y.; Hong, J.; Zhang, M.; Duan, W.; Mao, N.; Wu, J.; Xu, H.; Dong, F.; Lin, F., *et al.* Growth of Large-Area 2d $MoS_{2(1-x)}Se_{2x}$ Semiconductor Alloys. *Adv. Mater.* **2014**, 26, 2648-2653.

14. Liu, K.-K.; Zhang, W.; Lee, Y.-H.; Lin, Y.-C.; Chang, M.-T.; Su, C.-Y.; Chang, C.-S.; Li, H.; Shi, Y.; Zhang, H., *et al.* Growth of Large-Area and Highly Crystalline $MoS_2$ Thin Layers on Insulating Substrates. *Nano Lett.* **2012**, 12, 1538-1544.

15. Wu, S.; Huang, C.; Aivazian, G.; Ross, J. S.; Cobden, D. H.; Xu, X. Vapor–Solid Growth of High Optical Quality $MoS_2$ Monolayers with near-Unity Valley Polarization. *ACS Nano* **2013**, 7, 2768-2772.

16. Najmaei, S.; Liu, Z.; Zhou, W.; Zou, X.; Shi, G.; Lei, S.; Yakobson, B. I.; Idrobo, J.-C.; Ajayan, P. M.; Lou, J. Vapour Phase Growth and Grain Boundary Structure of Molybdenum Disulphide Atomic Layers. *Nat. Mater.* **2013**, 12, 754-759.

17. van der Zande, A. M.; Huang, P. Y.; Chenet, D. A.; Berkelbach, T. C.; You, Y.; Lee, G.-H.; Heinz, T. F.; Reichman, D. R.; Muller, D. A.; Hone, J. C. Grains and Grain Boundaries in Highly Crystalline Monolayer Molybdenum Disulphide. *Nat. Mater.* **2013**, 12, 554-561.

18. Lee, Y.-H.; Zhang, X.-Q.; Zhang, W.; Chang, M.-T.; Lin, C.-T.; Chang, K.-D.; Yu, Y.-C.; Wang, J. T.-W.; Chang, C.-S.; Li, L.-J., *et al.* Synthesis of Large-Area $MoS_2$ Atomic Layers with Chemical Vapor Deposition. *Adv. Mater.* **2012**, 24, 2320-2325.

19. Kaasbjerg, K.; Thygesen, K. S.; Jacobsen, K. W. Phonon-Limited Mobility in N-Type Single-Layer $MoS_2$ from First Principles. *Phys. Rev. B* **2012**, 85, 115317.

20. Schmidt, H.; Wang, S.; Chu, L.; Toh, M.; Kumar, R.; Zhao, W.; Castro Neto, A. H.; Martin, J.; Adam, S.; Özyilmaz, B., *et al.* Transport Properties of Monolayer $MoS_2$ Grown by Chemical Vapor Deposition. *Nano Lett.* **2014**, 14, 1909-1913.

21. Zhu, W.; Low, T.; Lee, Y.-H.; Wang, H.; Farmer, D. B.; Kong, J.; Xia, F.; Avouris, P. Electronic Transport and Device Prospects of Monolayer Molybdenum Disulphide Grown by Chemical Vapour Deposition. *Nat. Commun.* **2014**, 5, 3087.

22. Wu, J.; Schmidt, H.; Amara, K. K.; Xu, X.; Eda, G.; Özyilmaz, B. Large Thermoelectricity Via Variable Range Hopping in Chemical Vapor Deposition Grown Single-Layer $MoS_2$. *Nano Lett.* **2014**, 14, 2730-2734.

23. Wang, Q. H.; Kalantar-Zadeh, K.; Kis, A.; Coleman, J. N.; Strano, M. S. Electronics and Optoelectronics of Two-Dimensional Transition Metal Dichalcogenides. *Nat. Nanotechnol.* **2012**, 7, 699-712.

24. Splendiani, A.; Sun, L.; Zhang, Y.; Li, T.; Kim, J.; Chim, C.-Y.; Galli, G.; Wang, F. Emerging Photoluminescence in Monolayer MoS$_2$. *Nano Lett.* **2010**, 10, 1271-1275.

25. Castellanos-Gomez, A.; Agraït, N.; Rubio-Bollinger, G. Optical Identification of Atomically Thin Dichalcogenide Crystals. *Appl. Phys. Lett.* **2010**, 96, 213116.

26. Brivio, J.; Alexander, D. T. L.; Kis, A. Ripples and Layers in Ultrathin MoS$_2$ Membranes. *Nano Lett.* **2011**, 11, 5148-5153.

27. Mak, K. F.; Lee, C.; Hone, J.; Shan, J.; Heinz, T. F. Atomically Thin MoS$_2$: A New Direct-Gap Semiconductor. *Phys. Rev. Lett.* **2010**, 105, 136805.

28. Pankove, J. I. *Optical Processes in Semiconductors*; Courier Dover Publications: New York, 2012.

29. Perera, M. M.; Lin, M.-W.; Chuang, H.-J.; Chamlagain, B. P.; Wang, C.; Tan, X.; Cheng, M. M.-C.; Tománek, D.; Zhou, Z. Improved Carrier Mobility in Few-Layer MoS$_2$ Field-Effect Transistors with Ionic-Liquid Gating. *ACS Nano* **2013**, 7, 4449–4458.

30. Chakraborty, B.; Bera, A.; Muthu, D. V. S.; Bhowmick, S.; Waghmare, U. V.; Sood, A. K. Symmetry-Dependent Phonon Renormalization in Monolayer MoS$_2$ Transistor. *Phys. Rev. B* **2012**, 85, 161403.

31. Jahan, F.; Smith, B. E. Investigation of Solar Selective and Microstructural Properties of Molybdenum Black Immersion Coatings on Cobalt Substrates. *J. Mater. Sci.* **1992**, 27, 625-636.

32. Stewart, T. B.; Fleischauer, P. D. Chemistry of Sputtered Molybdenum Disulfide Films. *Inorg. Chem.* **1982**, 21, 2426-2431.

33. Anwar, M.; Hogarth, C. A.; Bulpett, R. Effect of Substrate Temperature and Film Thickness on the Surface Structure of Some Thin Amorphous Films of MoO$_3$ Studied by X-Ray Photoelectron Spectroscopy (ESCA). *J. Mater. Sci.* **1989**, 24, 3087-3090.

34. Baker, M. A.; Gilmore, R.; Lenardi, C.; Gissler, W. XPS Investigation of Preferential Sputtering of S from MoS$_2$ and Determination of MoSx Stoichiometry from Mo and S Peak Positions. *Appl. Surf. Sci.* **1999**, 150, 255-262.

35. Quan, M.; Patrick, M. O.; John, M.; Duy, L.; Chen, S. W.; Yeming, Z.; Tianyang, C.; Dezheng, S.; Koichi, Y.; Tai, T., *et al.* Controlled Argon Beam-Induced Desulfurization of Monolayer Molybdenum Disulfide. *J. Phys.: Condens. Matter* **2013**, 25, 252201.

36. Baugher, B.; Churchill, H. O.; Yang, Y.; Jarillo-Herrero, P. Intrinsic Electronic Transport Properties of High Quality Monolayer and Bilayer MoS$_2$. *Nano Lett.* **2013**, 13, 4212-4216.

37. Liu, H.; Neal, A. T.; Ye, P. D. Channel Length Scaling of MoS$_2$ MOSFETS. *ACS Nano* **2012**, 6, 8563-8569.

38.	Radisavljevic, B.; Kis, A. Mobility Engineering and a Metal–Insulator Transition in Monolayer MoS$_2$. *Nat. Mater.* **2013**, 12, 815-820.

39.	Ma, N.; Jena, D. Charge Scattering and Mobility in Atomically Thin Semiconductors. *Phys. Rev. X* **2014**, 4, 011043.

40.	Ghatak, S.; Pal, A. N.; Ghosh, A. Nature of Electronic States in Atomically Thin MoS$_2$ Field-Effect Transistors. *ACS Nano* **2011**, 5, 7707-7712.

41.	Qiu, H.; Xu, T.; Wang, Z.; Ren, W.; Nan, H.; Ni, Z.; Chen, Q.; Yuan, S.; Miao, F.; Song, F.*, et al.* Hopping Transport through Defect-Induced Localized States in Molybdenum Disulphide. *Nat. Commun.* **2013**, 4, 2642.

42.	Kim, B. H.; Park, M.; Lee, M.; Baek, S. J.; Jeong, H. Y.; Choi, M.; Chang, S. J.; Hong, W. G.; Kim, T. K.; Moon, H. R.*, et al.* Effect of Sulphur Vacancy on Geometric and Electronic Structure of MoS$_2$ Induced by Molecular Hydrogen Treatment at Room Temperature. *RSC Adv.* **2013**, 3, 18424-18429.

43.	Islam, M. R.; Kang, N.; Bhanu, U.; Paudel, H.; Erementchouk, M.; Tetard, L.; Leuenberger, M.; Khondaker, S. I. Electrical Property Tuning Via Defect Engineering of Single Layer MoS$_2$ by Oxygen Plasma. *Nanoscale* **2014**, 6, 10033-10039.

# Supporting Information

# Influence of Stoichiometry on the Optical and Electrical Properties of Chemical Vapor Deposition Derived MoS$_2$

In Soo Kim[1], Vinod K. Sangwan[1], Deep Jariwala[1], Joshua D. Wood[1], Spencer Park[1], Kan-Sheng Chen[1], Fengyuan Shi[1], Francisco Ruiz-Zepeda[5], Arturo Ponce[5], Miguel Jose-Yacaman[5], Vinayak P. Dravid[1,4], Tobin J. Marks[1,2], Mark C. Hersam[1,2,3,*], and Lincoln J. Lauhon[1,*]

[1]Department of Materials Science and Engineering, Northwestern University, Evanston, IL 60208 USA
[2]Department of Chemistry, Northwestern University, Evanston, IL 60208 USA
[3]Department of Medicine, Northwestern University, Evanston, IL 60208 USA
[4]International Institute of Nanotechnology, Northwestern University, Evanston, IL 60208 USA
[5]Department of Physics and Astronomy, University of Texas at San Antonio, San Antonio, TX 78249 USA

**Section 1: AFM images of MoS$_2$ flakes**

Grain boundaries in monolayer MoS$_2$ flakes were observed in both AFM topography and phase images. Figure S1a and b are low and high magnification topography images of Figure 1c. The images are taken from a region containing two merged monolayer flakes with smaller bilayers on top. In the low magnification image, bilayer regions are distinguished from monolayers by the contrast changes due to the height difference. Grain boundaries are observed in both mono- and bilayer flakes (marked by arrows). Figure S1c and d are low and high magnification phase images of Figure 1c. In both low and high magnification images, the contrast of grain boundaries is significantly improved. This suggests that grain boundaries can be readily imaged by AFM.

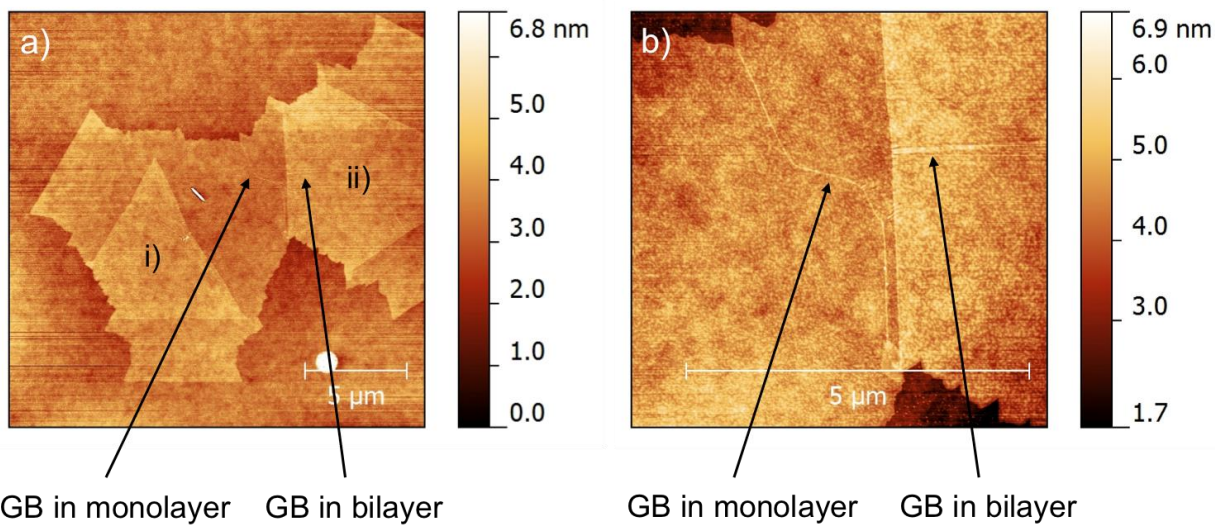

Figure S1. a) Low and b) high magnification AFM topography images.

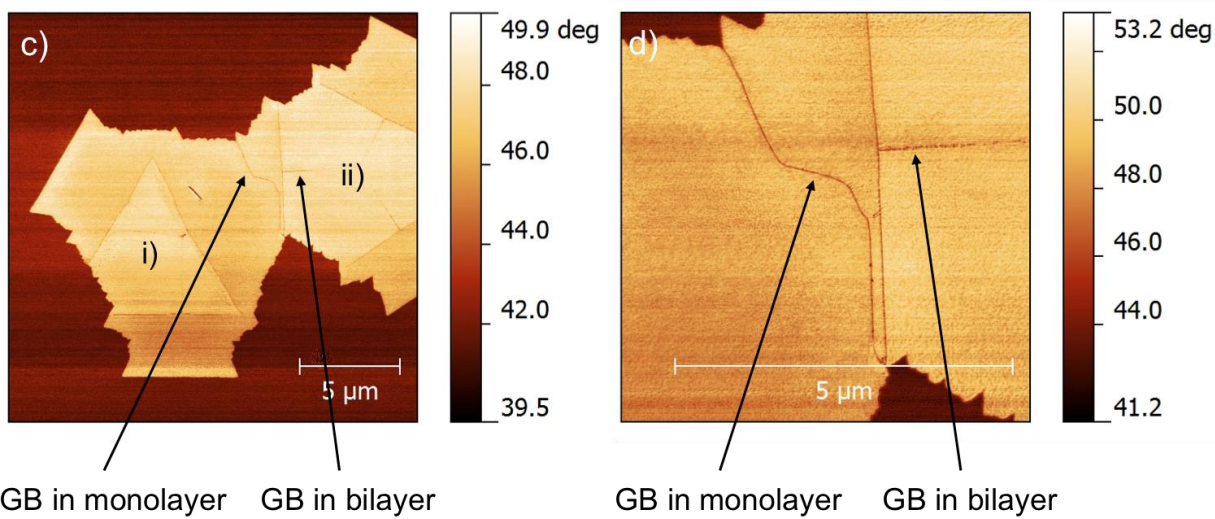

Figure S1. c) Low and d) high magnification AFM phase images.

**Section 2: Device characteristics and current-annealing group A**

As-fabricated group A devices showed relatively low conductance and field-effect mobility (0.005 – 0.1 cm$^2$/Vs). The devices also showed non-linear output curves at low biases (Fig S2a). However, the conductance was improved by biasing the devices at moderate biases $V_d$ = 5 V. Fig. S2b shows a real-time trace of the conductance of the device (Fig. S2a) as the gate bias is swept from 0 V → 60 V → -60 V → 0 V. Upon such 'current-annealing', the drain current at $V_g$ = 0 V increased by 20 times. Current annealing at $V_d$ = 5 V was repeated several times until only marginal increases in conductance were observed. Fig. S2c shows output characteristics of the device at the end of current-annealing cycles. The conductance has increased by an order of magnitude and the output curves have become more symmetric and linear. Fig. S2d and S3 show transfer curves of the same device in as-fabricated and final conditions. Although threshold voltage ($V_{th}$) measured at high drain bias ($V_d$ = 5 V) has shifted by -70 V (Fig S2b), the threshold voltage measured at low bias ($V_d$ = 0.5 V) has shifted by only 5 V (Fig. S3). In general, the field-effect mobility of the group-A devices could be increased by an order magnitude (range 0.1 – 4 cm$^2$/Vs) by current-annealing (Table T1).

Improvements in conductivity and mobility following current-annealing could be due to desorption of adsorbates by Joule heating. For example, absorbed oxygen molecules or S atoms may deplete the channel, increase energetic disorder, or act as carrier scattering centers. Finally, Joule heating may have favorable impact in the quality of contacts to MoS$_2$. Similar current-annealing strategies have been employed to improve mobility of graphene devices.[44] Recently, the performance of CVD MoS$_2$ transistors has also been improved by thermal annealing.[45] We also note that the effect of current-annealing on device performance was most clear in group A MoS$_2$ flakes. Group B and C MoS$_2$ flakes showed minimal irreversible changes in conductance at biases up to $V_d = 5$ V. We did not probe devices at higher biases due to increased tendency of devices to irreversibly degrade at higher fields.

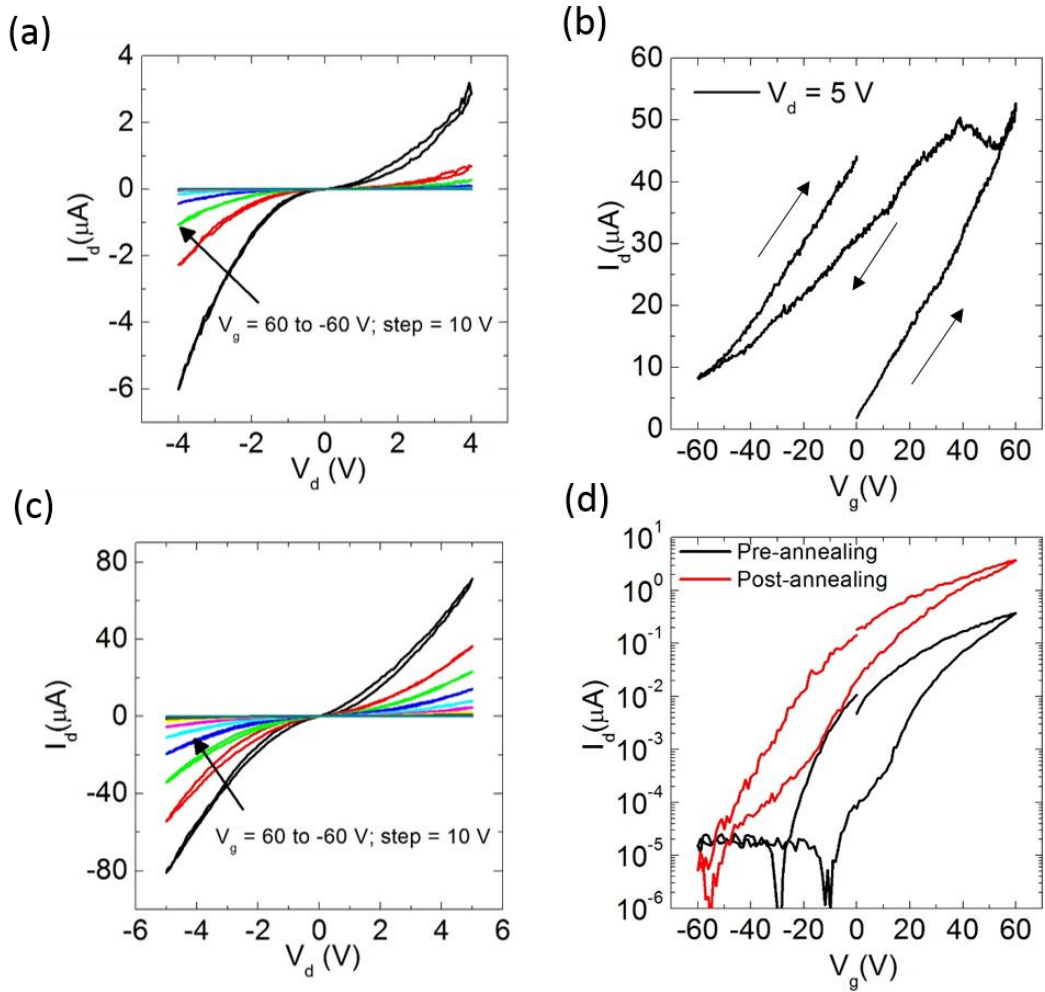

Figure S2: a) Output characteristics of an as-fabricated device (108-S2, Table 1) on group A MoS$_2$ flake. b) $I_d$ *versus* $V_g$ of the device at a large drain bias $V_d = 5$V during current annealing in vacuum (<2 x 10$^{-5}$ Torr). c) Output characteristics of the device after current-

annealing. d) Transfer characteristics of the device before and after current-annealing at $V_d$ = 0.5 V.

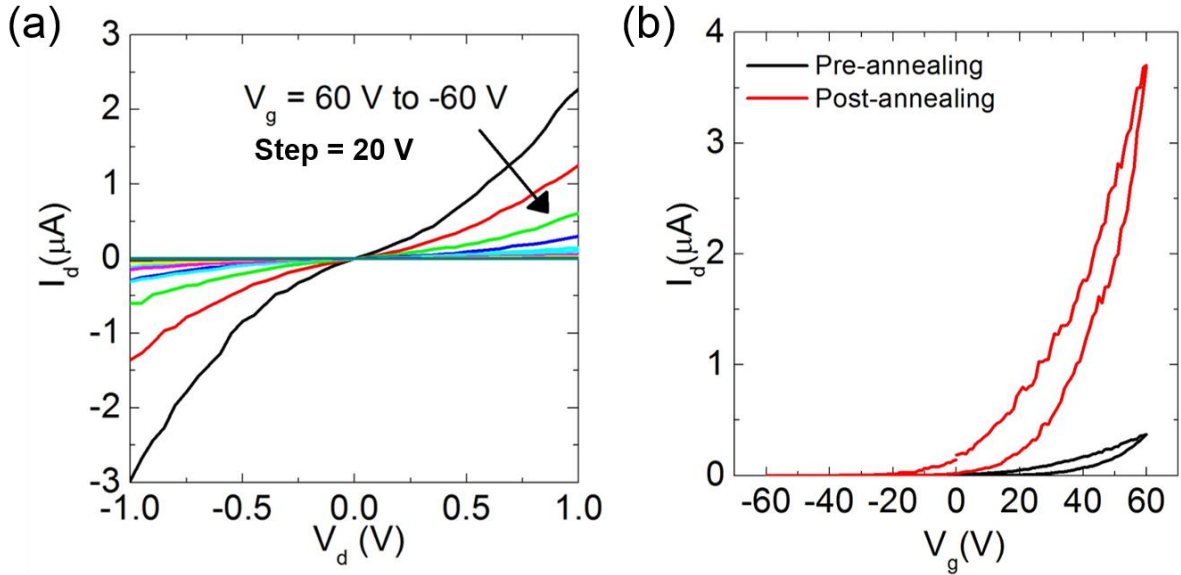

Figure S3: a) Output characteristics data from Fig. S2c at low biases. B) Transfer characteristics data from Fig. S2d shown in linear plot.

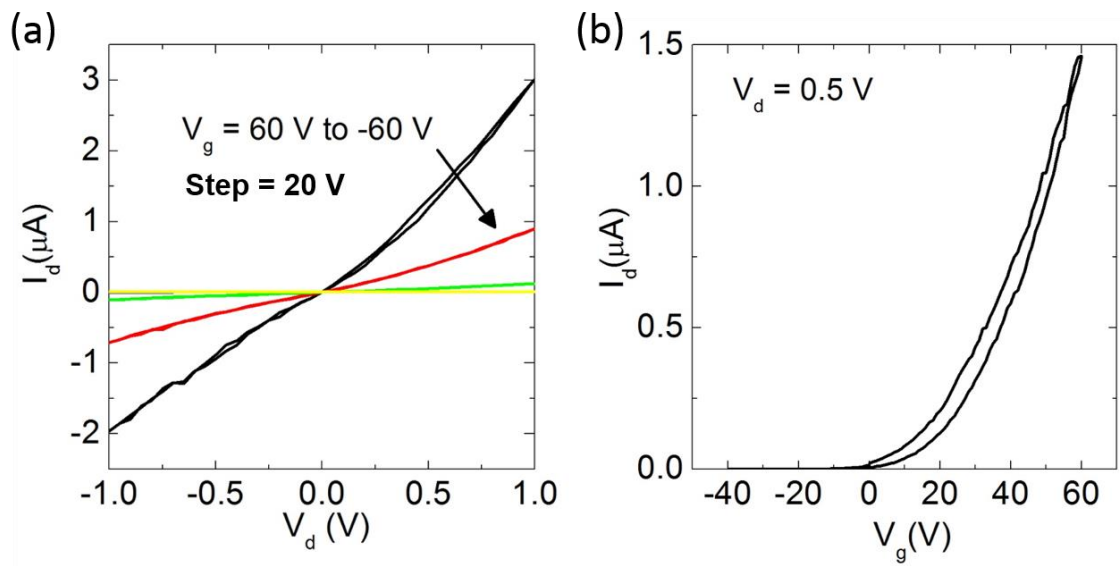

Figure S4: a) Output characteristics, and b) transfer characteristics of a group B device (78-S2, Table 1).

Table T1. Channel dimensions, field-effect mobility, and threshold voltage of group A, B, and C devices investigated.

| Growth Details | Sample Number | Channel Dimensions L X W (µm) | Mobility (cm²/Vs) | Threshold Voltage (V) |
|---|---|---|---|---|
| Minimum Sulfurization (Group-1) | 68-S2 | 4 x 3 | 18.2 | 13 |
| | 68-S3 | 0.5 x 1.5 | 13.6 | 10 |
| | 68-S4 | 0.5 x 2 | 12.5 | 13 |
| | 68-S5 | 0.5 x 8 | 12.3 | 6 |
| | 68-S6 | 4 x 5 | 21.8 | 10 |
| | 68-S4n | 3 x 4 (4p)** | 13.6 | 22 |
| Medium Sulfurization (Group-2) | 78-S1 | 2 x 5 | 3.6 | 32 |
| | 78-S2 | 2 x 5.3 | 2.9 | 30 |
| | 74-S1 | 8 x 6 | 3.6 | 43 |
| | 74-S2 | 8.5 x 4 | 0.2 | 28 |
| | 73-S1 | 8.5 x 10 | 3.9 | 48 |
| | 73-S2 | 14 x 14.5 | 1.8 | 27 |
| Maximum Sulfurization (Group-3) | 108-S1 | 4 x 6.5 | 1.1* | 43 |
| | 108-S2 | 0.5 x 7 | 0.04/0.45* | 39 |
| | 108-S3 | 4 x 7.8 | 1.5* | 44 |
| | 108-S4 | 0.5 x 10.5 | 0.07/0.8* | 38 |
| | 114-S1 | 2 x 6 (4p)** | 0.006/0.16* | 47 |
| | 114-S3 | 2 x 6 (4p)** | 0.13/3.9* | 34 |

* denotes field-effect mobility values after current annealing (only in Group 3 samples).

** denotes devices measured in four-terminal geometry.

For devices showing large hysteresis, the field-effect mobility values were obtained by averaging slopes of the two curves.

**References**

1. Bolotin, K. I.; Sikes, K. J.; Jiang, Z.; Klima, M.; Fudenberg, G.; Hone, J.; Kim, P.; Stormer, H. L. Ultrahigh Electron Mobility in Suspended Graphene. Solid State Commun. 2008, 146, 351-355.

2. Schmidt, H.; Wang, S.; Chu, L.; Toh, M.; Kumar, R.; Zhao, W.; Castro Neto, A. H.; Martin, J.; Adam, S.; Özyilmaz, B., et al. Transport Properties of Monolayer $MoS_2$ Grown by Chemical Vapor Deposition. Nano Letters 2014, 14, 1909-1913.